\documentclass[aps,pre,reprint,floatfix]{revtex4-1}	
\usepackage[utf8]{inputenc}     
\usepackage[english]{babel}  

\usepackage{bm}
\usepackage{amsmath,amsfonts,amssymb}
\usepackage[hidelinks,colorlinks=true]{hyperref}

\usepackage{xcolor}
\usepackage{graphicx}

\definecolor{myred}{RGB}{228,26,28}
\definecolor{myblue}{RGB}{55,126,184}
\definecolor{myorange}{RGB}{225,127,0}
\definecolor{mygreen}{RGB}{77,175,74}
\definecolor{mylila}{RGB}{152,78,163}
\definecolor{mybrown}{RGB}{153,76,0}
\definecolor{mygray}{RGB}{153,153,153}
\definecolor{darkred}{rgb}{0.8,0,0}
\definecolor{mydarkgreen}{RGB}{0,102,0}
\definecolor{mydarkbrown}{RGB}{102,52,0}
\definecolor{Orange}{RGB}{235,129,27}
\definecolor{Green}{RGB}{35,55,59}
\usepackage{tikz} 
\usetikzlibrary{decorations,backgrounds,patterns} 
\usepackage{pgfplots} 
\usepackage{pgfplotstable}
\usepgfplotslibrary{external}
\pgfplotsset{compat=newest}
\usetikzlibrary{pgfplots.groupplots}
\usetikzlibrary{calc}
\usetikzlibrary{3d}
\usetikzlibrary{positioning}
\usepgfplotslibrary{units}
\pgfplotsset{unit code/.code={\si{#1}}}
\usepgfplotslibrary{patchplots}
\usepgfplotslibrary{fillbetween}
\tikzset{>=stealth}

\renewcommand{\d}{\mathrm{d}}

\newcommand{\euler}{\mathrm{e}}

\newcommand{\energy}{E}

\newcommand{\chempot}{\mu}
\newcommand{\invtemperature}{\beta}
\newcommand{\efficiency}{\eta}

\newcommand{\power}{\mathcal{P}}

\newcommand{\single}{s}
\newcommand{\majority}{m}
\newcommand{\change}{\Delta}

\newcommand{\dimension}{N}
\newcommand{\threshold}{\zeta}

\newcommand{\microprobability}{p}

\newcommand{\macroprobability}{P}

\newcommand{\microrates}{\omega}

\newcommand{\microenergy}{e}
\newcommand{\microheat}{q}
\newcommand{\microwork}{w}
\newcommand{\microentropy}{s}

\newcommand{\macroenergy}{E}
\newcommand{\macroheat}{Q}
\newcommand{\macrowork}{W}
\newcommand{\macroentropy}{S}

\newcommand{\microefficiency}{\eta_{\single}}
\newcommand{\macroefficiency}{\eta_{\majority}}
\newcommand{\micropower}{\mathcal{P}_{\single}}
\newcommand{\macropower}{\mathcal{P}_{\majority}}
\newcommand{\betafunction}{\mathcal{I}}

\begin{document}

\title{Thermodynamics of majority-logic decoding in information erasure}

\author{Shiqi Sheng$^1$, Tim Herpich$^{2}$, Giovanni Diana$^{3}$, and Massimiliano Esposito$^{2}$}

\affiliation{$^1$Division of Interfacial Water and Key Laboratory of Interfacial Physics and Technology, Shanghai Institute of Applied Physics, Chinese Academy of Sciences, Shanghai 201800, China}
\affiliation{$^2$Complex Systems and Statistical Mechanics, Physics and Materials Science Research Unit, University of Luxembourg, L-1511 Luxembourg, Luxembourg}
\affiliation{$^3$Centre for Developmental Neurobiology and MRC Center for Neurodevelopmental Disorders, King's College London, Guy's Hospital Campus London SE1 1UL, United Kingdom}
\email{Electronic Mail: massimiliano.esposito@uni.lu}
\date{\today}

\begin{abstract}
We investigate the performance of majority-logic decoding in both reversible and finite-time information erasure processes performed on macroscopic bits that contain $\dimension$ microscopic binary units. While we show that for reversible erasure protocols single-unit transformations are more efficient than majority-logic decoding, the latter is found to offer several benefits for finite-time erasure processes:
Both the minimal erasure duration for a given erasure and the minimal erasure error for a given erasure duration are reduced, if compared to a single unit. Remarkably, the majority-logic decoding is also more efficient in both the small erasure error and fast erasure region. These benefits are also preserved under the optimal erasure protocol that minimizes the dissipated heat. Our work therefore shows that majority-logic decoding can lift the precision-speed-efficiency trade-off in information erasure processes.
\end{abstract}

\maketitle

\onecolumngrid

\section{Introduction} \label{sec:introduction}
For modern society that is characterized by ubiquitous digitalization, information storage and processing are of utmost importance. Since it becomes increasingly challenging to keep the capacity of the ever-smaller storage media constant, the speed and efficiency of information processing and thus the underlying thermodynamics of the storage medium are becoming more crucial.
Moreover, the exceedingly large thermodynamic costs of performing computation poses a tremendous technological challenge.
Researchers have addressed these technological challenges by studying the precision-speed-efficiency trade-off inherent to bit erasure---arguably, the simplest type of computation---using recent developments from nonequilibrium statistical mechanics.

The relationship between thermodynamics and information theory traces back to Maxwell who proposed the notorious Maxwell demon \cite{MaxwellDemon2002}. Later, Szil\'{a}rd derived a relationship between the information consumed by the demon and the extractable work from a thermal reservoir \cite{Szilard1929}.
Shannon established the foundations of information theory by his seminal works \cite{Shannon1948BSTJ1,Shannon1948BSTJ2}, in which information is interpreted as the uncertainty of an outcome of an event and the so-called Shannon entropy is introduced to quantify the amount of information.
Based on these works, a milestone in understanding the connection between thermodynamics and information processing was achieved by the so-called Landauer's principle which was established by the work from Landauer \cite{Landauer1961IBM}, Bennett~\cite{Bennett1973IBM,Bennett1982} and Penrose~\cite{Penrose1970}.
Landauer's principle provides an explicit lower bound for the heat dissipated by the system when one bit of information is erased. The validity of this lower bound has been verified experimentally \cite{Berut2012,Berut2015,Jun2014} and reproduced in more general theoretical works such as \cite{Piechocinska2000,Shizume1995,Esposito2011EPL}.
The lower bound provided by Landauer's principle is only achieved in quasi-static processes, i.e., for information erasing processes of infinite duration.

For practical purposes one is naturally interested in fast-erasure processes, hence
limiting the applicability of Landauer's principle and prompting multiple works on finite-time information erasure and other elementary logical operations \cite{owen2017arxiv,wolpert2017arxiv,Kolchinsky2017IF,wolpert2018arxiv,Sagawa2017SPRINGER,parrondo2015nature}.
A novel theory referred to as stochastic thermodynamics \cite{Seifert2012RPP,broeck2015physica}, allows to systematically address the thermodynamic properties in out-of-equilibrium systems. Owing to the development of stochastic thermodynamics and to the aforementioned work of Shannon, novel methods to study finite-time erasure processes became available and prompted several works, e.g., \cite{Esposito2010EPL,Giovanni2013PRE,Zulkowski2014PRE,Gavrilov2017PNAS}.
This fundamental understanding of the thermodynamic costs in finite-time bit erasure may provide new computing strategies for the information and technology industry.

As a universal result, it was, for instance, proven that in the low-dissipation limit the optimal, i.e., the least work-intense, transformation protocol between two sets of probabilities will lead to an irreversible entropy production, i.e., dissipation, that is inversely proportional to the transformation duration \cite{Aurell2012JSP,Schmiedl2008EPL,zulkowski2015pre}. This result has also been applied beyond the theory of information processing, e.g., for the study of efficiencies of finite-time heat engines \cite{Esposito2010PRL,Schmiedl2007PRL,TuSheng2014PRE,TuSheng2015NJP,TuSheng2015PRE,Esposito2010PRE}. 
However, away from the low-dissipation regime, the relationship between irreversible entropy production and operating duration is model-dependent.
While two-state systems with Fermi transition rates as a representation of a bit have been investigated \cite{Esposito2010EPL,Giovanni2013PRE}, their applicability to real systems is limited.
In general, the correspondence between stored information and a two-state unit is not a one-to-one, but one bit of information is normally stored in a macroscopic bit that, in turn, is composed of an array of non-interacting microscopic binary units.
As an example, in magnetic storage media one bit of information is stored in a macroscopic bit that contains tens to hundreds of microscopic magnetic grains.
How the collective physical information stored in the array of bistable microscopic units is translated into macroscopic logical information stored by the bit is determined by specific decoding rules. The decoding procedure can formally be understood as a coarse-graining of microscopic physical information. Employing decoding procedures, the safety of the information processing is enhanced since the signal-to-noise ratio is proportional to $\sqrt{\dimension}$, where $\dimension$ is the number of microscopic grains contained in a macroscopic bit \cite{Richter1999}.

It is the aim of this paper to extend the existing studies of the thermodynamics of finite-time information erasure processes in single microscopic two-state systems by considering non-interacting ensembles of them under a specific decoding procedure.
To the best of our knowledge, there is no systematic discussions on the energetics of information erasure processes under coarse-grained decoding in the literature. Hence, in this work we explore the costs and benefits of employing the majority-logic decoding, which is the simplest and most used coarse-graining method in information processing \cite{Majoritylogicdecoding}.
We show that for reversible erasure protocols, information erasure in single units is more efficient than symmetric majority-logic decoding. Conversely, we find that for finite-time erasure protocols the majority-logic decoding can accelerate the process of erasing a fixed amount of information or compress the minimal erasure error of a fixed-time erasure process. While these benefits in terms of speed and precision for most erasure processes come at the expense of a lower erasure efficiency, we show that, remarkably, the majority-logic decoding will however also be more efficient than a single-unit process when the erasing is fast, or the erasure error is small. 
When imposing the optimal erasure protocol that minimizes the dissipated heat, we find that for the two unit models investigated in this work (Fermi- and Arrhenius-rates units), these advantages are preserved. Hence, we conclude that the majority-logic decoding lifts the trade-off between erasure speed, precision, and efficiency when compared to a single unit.

The plan of the paper is as follows: In Section \ref{sec:majorityrule} we introduce the microscopic storage unit and define two types of macroscopic bits, the single-unit bit and the majority-logic decoding bit. The latter is mathematically formulated.
Section \ref{sec:finitetimeprocesses} is devoted to studying the energetics of information erasure processes with majority-logic decoding for different protocols: First, a constant state-energy protocol of the microscopic unit is considered. Next, two microscopic unit models, the Fermi-rates and the Arrhenius-rates model, are introduced and studied if the optimal protocol for the erasure process is imposed on the system. In the high- and low-dissipation limit analytic results are derived. Finally, we conclude with a summary and an outlook to potential future projects in Section \ref{sec:conclusion}.

\section{Majority-Logic Decoding\label{sec:majorityrule}}
\unskip
\subsection{Two Setups of Macroscopic Bits}
\begin{figure}[h!]
\begin{center}

\includegraphics[scale=1]{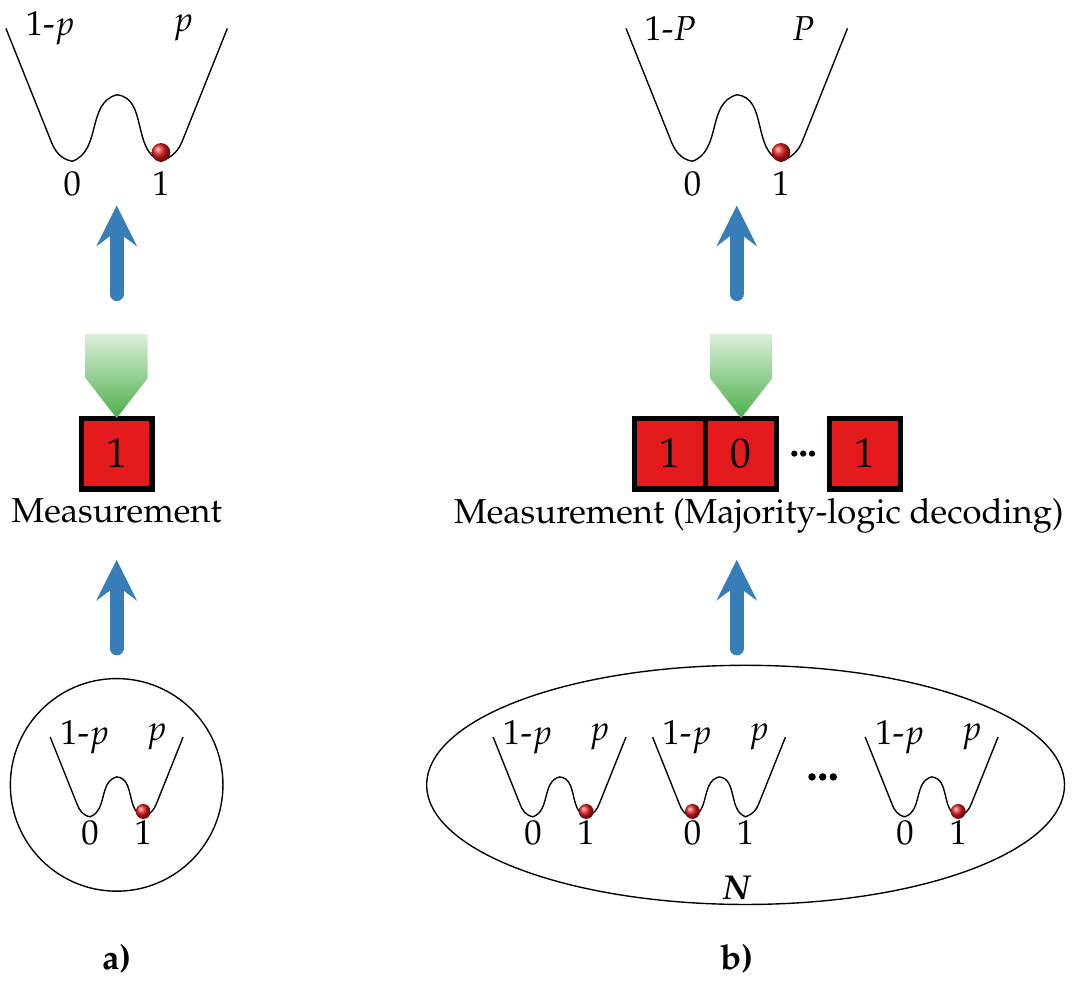}

\caption{Schematics of a single-unit bit (SUB) in (\textbf{a}) and a $\dimension$-majority-logic decoding bit (MLB) in (\textbf{b}).  \label{fig:schematics} }
\end{center}
\end{figure}

As an elementary storage unit, we consider a microscopic binary unit with states $0$ and $1$ and denote by $\microprobability$ the probability to observe it in state $1$. Based on these microscopic binary units we can construct macroscopic (logical) bits in two different ways as illustrated in Figure \ref{fig:schematics}.
First, we consider a single-unit bit (SUB) in Figure \ref{fig:schematics}a that consists of only one microscopic unit and is in contact with a heat bath at inverse temperature $\invtemperature$. The probability of finding the SUB in state $1$ is represented by $\macroprobability$, and is, of course, equal to probability $\microprobability$ for the microscopic unit to be in state $1$. Alternatively, a majority-logic decoding bit (MLB) can be thought of as an array of $\dimension$ identical and non-interacting microscopic units that are subjected to the same experimental protocol and connected to a heat bath at inverse temperature $\invtemperature$ as sketched in Figure \ref{fig:schematics}b.
The probability of the MLB to be in state $1$ is denoted by $\macroprobability$ and is determined via majority-logic decoding. This decoding scheme prescribes that the information encoded in the MLB corresponds to the state that is occupied most at the level of the microscopic units and is therefore a coarse-graining procedure that is mathematically formulated in the following.

\subsection{From Microscopic to Macroscopic Probability}
\begin{figure}[h!]
\begin{center}

%
%
%

\includegraphics[scale=1]{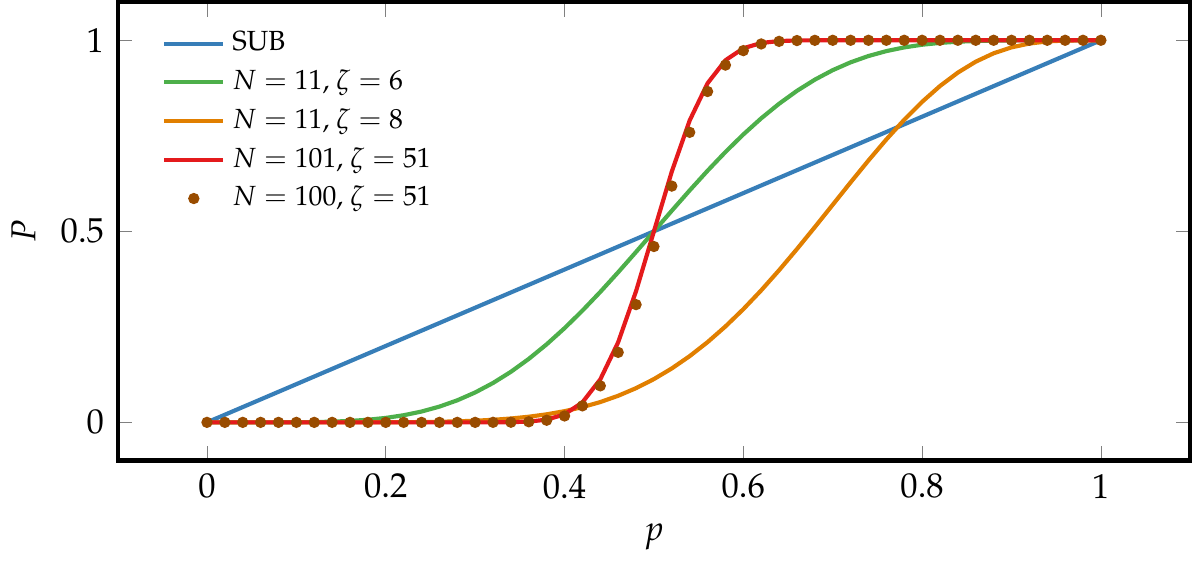}

\caption{The curves of $\macroprobability$ with respect to $\microprobability$ at different values of $\dimension$ and $\threshold$.  The blue line refers to the SUB. While the green ($\dimension=11,\threshold=6$) and red curves ($\dimension=101,\threshold=51$) refer to symmetric MLBs, the orange ($\dimension=11,\threshold=8$) and the brown dotted curves ($\dimension=100,\threshold=51$) refer to asymmetric MLBs with even $\dimension$. \label{fig:micromacroprobability} }
\end{center}
\end{figure}

We proceed by formulating the relation between the microscopic probability $\microprobability$ associated with each microscopic unit and the macroscopic probability $\macroprobability$ specifying the occupation probability of the MLB. This relation can be expressed as
\begin{align} \label{eq:majorityruledecoding}
\macroprobability(\microprobability,\dimension,\threshold)=\sum_{k=\threshold}^{\dimension}{\binom \dimension k} \microprobability^{k} \, (1-\microprobability)^{\dimension-k} ,
\end{align}
where $\threshold$ represents the threshold number of the detector of the same microscopic state during the measurement process. In Appendix \ref{sec:proofbinomialbetaequality} the equality
\begin{align}
\macroprobability(\microprobability,\dimension,\threshold)=\betafunction_{\microprobability}(\threshold,N+1-\threshold).\label{eq:binomialbetaequality}
\end{align}
is proven. Here, $\betafunction$ refers to the regularized incomplete beta function
\begin{align}
\betafunction_{x}(a,b)=\frac{\int_{0}^{x}t^{a-1}(1-t)^{b-1}\mathrm{d}t}{\int_{0}^{1}t^{a-1}(1-t)^{b-1}\mathrm{d}t} \label{eq:betafunction}.
\end{align}
The bijectivity of the regularized incomplete beta function $\betafunction$ allows furthermore to determine the microscopic probability $\microprobability$ given the macroscopic probability $\macroprobability$ of an MLB as follows
\begin{align}
\microprobability(\macroprobability,\dimension,\threshold)=\betafunction_{\macroprobability}^{-1}(\threshold,\dimension+1-\threshold).  \label{eq:binomialbetainverse}
\end{align}
The qualitative features of the majority-logic decoding are illustrated in Figure \ref{fig:micromacroprobability} that compares the macroscopic probability $ \macroprobability(\microprobability,\dimension,\threshold) $ of a SUB and MLB for different array sizes $\dimension$ and threshold values~$\threshold$.

First, we observe that the macroscopic probability $\macroprobability$ is monotonically increasing with $\microprobability$. Next, for any value of $\dimension$ and $\threshold$ it holds for $x=0,1$ that $\macroprobability(x,\dimension,\threshold)=x$ since the complete beta function becomes a unity operator, $\betafunction = \bm{I}$. Physically, this means that perfect information erasure in an MLB ($\macroprobability =0,1$) is realized by perfect erasure in each microscopic unit the MLB consists of.
For odd values of $\dimension$ the symmetric case, $\threshold=(\dimension+1)/2$, corresponds to the \emph{majority}-logic decoding in the strict sense, that is
\begin{align} \label{eq:majorityruledecodingsymmetric}
\macroprobability \left(\microprobability,\dimension, \tfrac{\dimension+1}{2} \right) = \betafunction_{\microprobability}\left( \tfrac{\dimension+1}{2},\tfrac{\dimension+1}{2}\right). 
\end{align}
We note that for symmetric majority-logic decoding, $\threshold=(\dimension+1)/2$, the curve is symmetric with respect to $(1/2,1/2) \; \forall \dimension$. This is readily derived by noting that according to the binomial theorem, one has
\begin{align}
\macroprobability\big(\tfrac{1}{2},\dimension,0 \big) = \frac{1}{2^{\dimension}} \sum_{k=0}^{\dimension}{\binom \dimension k}  = \frac{1}{2^{\dimension}} (1+1)^{\dimension}  = 1.
\end{align}
Thus, it immediately follows from the symmetry of the binomial probability distribution (Equation~ \eqref{eq:majorityruledecoding}) that one has $ \macroprobability\big(1/2,\dimension,\threshold = (\dimension+1)/2 \big) = 1/2 $.
It can furthermore be seen in Figure \ref{fig:micromacroprobability} that for large $\dimension$ the curves converge to a step function centered at $\microprobability = 1/2$. For large $\dimension$, Taylor-expanding Equation \eqref{eq:majorityruledecoding} for the symmetric case around $\microprobability=1/2$ up to linear order in $| \microprobability - 1/2 |$ yields
\begin{align}
\left| \frac{1}{2} - \macroprobability \right| = \sqrt{\frac{2 \dimension}{\pi}} \left| \frac{1}{2} - \microprobability \right|  + \mathcal{O} \bigg( \left| \frac{1}{2}-\microprobability \right|^{3} \bigg) \, ,  \label{eq:micromacroprobabilityapproximation}
\end{align}
where we denote by $\mathcal{O}(x^2)$ all terms that are at the order $x^2$ or higher. According to Equation \eqref{eq:micromacroprobabilityapproximation}, the slope at the symmetric point $\microprobability$ is increasing with $ \sqrt{\dimension} $ in the symmetric decoding case. We therefore arrive at the first important result that large macroscopic erasure in the MLB can be achieved at the cost of small microscopic erasure in each microscopic unit.
The good agreement of the curves corresponding to $\dimension=100$ and $\dimension=101$ suggests that for $\threshold \approx (\dimension+1)/2$ the symmetry decoding case is approached if $\dimension$ becomes large.

\subsection{Majority-Logic Decoding as a Coarse-Graining Procedure}
\label{sec:coarsegraining}

The equivalent probabilistic descriptions on the microscopic and macroscopic level prompt the question of how to define the underlying physical processes at these levels. In information erasure processes, important quantities are the heat that is generated during the operation, the change in Shannon entropy that measures the amount of information erased by that process as well as the erasure efficiency \cite{Esposito2011EPL,Giovanni2013PRE}.

On the level of a single microscopic unit, the heat dissipated during an erasure process from $\microprobability_i$ to $\microprobability_f$ is denoted by $\microheat$ and defined as a negative quantity $\microheat < 0$. With this convention, the first law of thermodynamics in differential form reads
\begin{align}
\d_t \, \microenergy = \dot{\microheat} + \dot{\microwork} ,
\end{align}
where $ \dot{\microwork}$ denotes the work current and $ \dot{\microenergy}$ the rate of energy change. The microscopic Shannon entropy
\begin{align} 
\microentropy(\microprobability)= - \microprobability \ln{\microprobability} - (1-\microprobability) \ln{(1-\microprobability)} ,
\label{eq:microentropy}
\end{align}
allows to quantify $\change \microentropy(\microprobability_i,\microprobability_f) \equiv \microentropy(\microprobability_{f}) - \microentropy(\microprobability_{i}) $ as the amount of information that is erased during an erasure process from $\microprobability_i$ to $\microprobability_f$. We set the Boltzmann constant $k_{b} \equiv 1$. The second law of thermodynamics reads
\begin{align}
\d_t \, \microentropy = \invtemperature \dot{\microheat} + \dot{\sigma} ,
\end{align}
where $\dot{\sigma} \geq 0$ refers to the irreversible entropy production rate in the microscopic unit.

Throughout this work, we consider the case in which the initial state is the maximum information state given by $\microprobability_i =1/2$ and $\microentropy(1/2) = \ln 2$. For this case, the change in entropy is always negative $\change \microentropy(\microprobability_i,\microprobability_f) < 0$ if $\microprobability_f \neq 1/2 $. Therefore, a suitable definition of the microscopic erasure efficiency for this process reads
\begin{align}
\microefficiency(\microprobability_i,\microprobability_f) = \frac{\change \microentropy(\microprobability_i,\microprobability_f)}{\invtemperature \, \microheat} .  \label{eq:microefficiency}
\end{align}
The heat $\microheat$ generated by the microscopic unit is naturally dependent on specific models and operating protocols. The optimal protocols that minimize the dissipated heat and thus maximize the erasure efficiency are investigated in Section \ref{sec:finitetimeprocesses}. 

On the level of macroscopic bits, the heat dissipation refers to the cumulated heat generated by all the microscopic units the bit consists of. Thus, the microscopic definitions from above are, of course, also physically significant for the erasure process at the level of the SUB that is $ \microentropy(\macroprobability) = \microentropy(\microprobability)$ and $\microefficiency(1/2,\microprobability_f) = \microefficiency(1/2,\macroprobability_f) $.
Conversely, to perform a macroscopic erasure in the MLB from $\macroprobability_i=1/2$ to $\macroprobability_f$, an amount of information specified by the majority-logic decoding needs to be erased in each microscopic unit contained in the MLB.
Here, $\macroprobability_f$ is the probability that the final state of the MLB after the erasure is logically decoded as state 1. According to Equation \eqref{eq:binomialbetainverse}, this amounts to change in each microscopic unit from the initial state $\microprobability_{i}=\betafunction^{-1}_{1/2}(\threshold,\dimension+1-\threshold) $ to the final state $\microprobability_{f}=\betafunction^{-1}_{\macroprobability_{f}}(\threshold,\dimension+1-\threshold)$.
The heat dissipated by the MLB, $\macroheat$, is thus determined as follows
\begin{align}
\macroheat(1/2 \to \macroprobability_f) = \dimension \, \microheat \left(  \betafunction^{-1}_{1/2}(\threshold,\dimension+1-\threshold) \to \betafunction^{-1}_{\macroprobability_{f}}(\threshold,\dimension+1-\threshold) \right) . \label{eq:macroheat}
\end{align}
The first law of thermodynamics at the level of the MLB reads
\begin{align}
\d_t \macroenergy = \dot{\macroheat} + \dot{\macrowork} ,
\end{align}
where the thermodynamic quantities of the MLB are naturally given by the sum of the microscopic ones, i.e., one has $ \macroenergy = \dimension \microenergy , \macroheat = \dimension \microheat , \macrowork = \dimension \microwork $.
It furthermore holds that $\Sigma = \dimension \sigma$ since the irreversible entropy production of the MLB must be equal to the sum of the irreversible entropy production of the microscopic units contained in the MLB.
Next, we define the entropy quantifying the information stored in the MLB as for the microscopic unit, i.e.,
\begin{align} \label{eq:macroentropy}
\macroentropy(\macroprobability ) = - \macroprobability \ln{\macroprobability} - (1-\macroprobability) \ln{(1-\macroprobability)} .
\end{align}
From the definition for the entropy associated with an MLB made in Equation \eqref{eq:macroentropy} follows that $\macroentropy \neq \dimension \microentropy $. This however implies that the entropy balance at the level of the MLB is broken \cite{herpich2018prx,Esposito2012PRE,herpich2019pre}
\begin{align}
\d_t \macroentropy \neq \invtemperature \, \dot{\macroheat} + \dot{\Sigma} .
\end{align}
Hence the macroscopic Shannon entropy in Equation \eqref{eq:macroentropy} should be thought of as logical but not strictly physical information.
Finally, the macroscopic efficiency associated with the erasure in the MLB is defined as
\begin{align}
\macroefficiency(1/2,\macroprobability_f) = \frac{ \change  \macroentropy(1/2,\macroprobability_f) }{ \invtemperature \, \macroheat },\label{eq:macroefficiency}
\end{align}
with the change in macroscopic Shannon entropy $ \change \macroentropy(1/2,\macroprobability_f) \equiv \macroentropy(\macroprobability_f) - \macroentropy(1/2) = \macroentropy(\macroprobability_f) - \ln 2 $.

\subsection{Reversible Erasure Protocols}
\label{sec:reversibleerasure}

In the limit of reversible erasure, the irreversible entropy production vanishes, and one has for the heat in a SUB $ \invtemperature \, \microheat = \microentropy(\microprobability_f) - \ln 2$. Thus, from Equation \eqref{eq:microefficiency} follows that $\microefficiency^{rev} = 1$ for any erasure process in this limit.
Turning to the MLB, the dissipated heat during the erasure process reads
\begin{align} 
\invtemperature \, \macroheat =  \dimension \left[ \microentropy \Big( \betafunction^{-1}_{\macroprobability_f}(\threshold,\dimension+1-\threshold)  \Big)  - 
\microentropy \Big( \betafunction^{-1}_{\tfrac{1}{2}}(\threshold,\dimension+1-\threshold)  \Big)
 \right] .
\end{align}

\begin{figure}[h!]
\begin{center}

\includegraphics[scale=1]{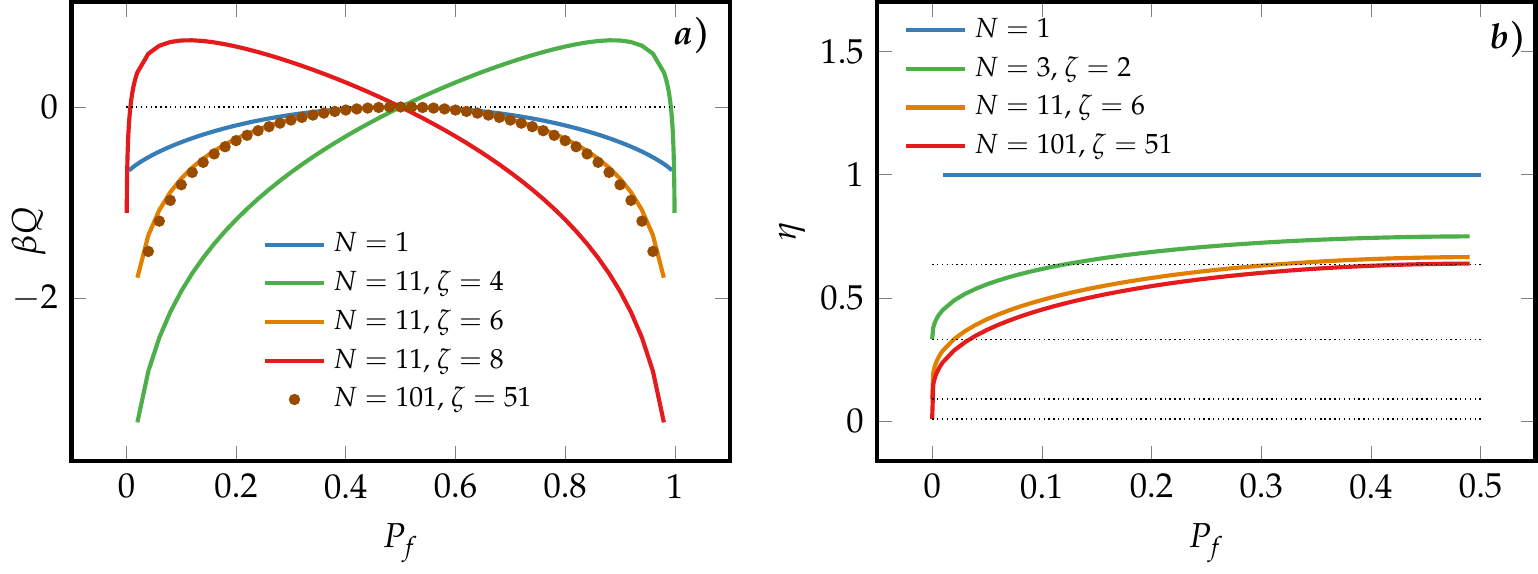}

\caption{(\textbf{a}) Heat dissipated by the macroscopic bits in the reversible limit. The initial macroscopic state is $\macroprobability_{i}=1/2$; (\textbf{b}) Efficiencies associated with reversible erasure processes for the SUB and symmetric MLBs. The dotted lines from top to bottom represent the values $\efficiency=2/\pi$, $\efficiency=1/3$, $\efficiency=1/11$ and $\efficiency=1/101$, respectively.\label{fig:heatandefficiencyreversible}}
\end{center}
\end{figure}

Figure \ref{fig:heatandefficiencyreversible} depicts in the reversible limit the heat dissipated during an erasure process from $\macroprobability_i = 1/2$ to $\macroprobability_f$ by a SUB and MLB [panel (a)] and the associated efficiencies for symmetric decoding processes [panel (b)].
We observe in panel (a) that the SUB always dissipates less heat than the symmetric MLBs which generate more heat as $\dimension$ increases. For asymmetric majority-logic decoding, $ \threshold \neq (\dimension+1)/2$, the heat generated by a MLB is reduced with respect to the one of a SUB and even takes positive values for probabilities in the range $[1/2,\betafunction_{1-\betafunction^{-1}_{\macroprobability_f}(\threshold,\dimension+1-\threshold)}(\threshold,\dimension+1-\threshold)]$ or $[\betafunction_{1-\betafunction^{-1}_{\macroprobability_f}(\threshold,\dimension+1-\threshold)}(\threshold,\dimension+1-\threshold),1/2]$, if the decoding is left-asymmetric or right-asymmetric, respectively.
The positivity of the heat means that one could even extract work during certain erasure processes by employing asymmetric majority-logic decoding. This property is similar to the division of logical entropy of the system entropy and the physical entropy of each subspaces as discussed in Refs. 
\cite{Cover1991,Sagawa2014JSM,Gavrilov2017PRL}. However, here, the analogue of physical entropy in each subspace varies according to Equations \eqref{eq:majorityruledecoding} and \eqref{eq:microentropy} as the macroscopic probability $\macroprobability$ changes.

The erasure efficiency of MLBs under symmetric majority-logic decoding, $ \threshold = (\dimension+1)/2$, for different array lengths $\dimension$ is shown in Figure \ref{fig:heatandefficiencyreversible}b. Consistently to the observations made in panel (b), the SUB is always more efficient than the MLB whose erasure efficiency decreases as $\dimension$ increases.
The same applies to the other erasure branch given by $\macroprobability_f > 1/2$ that can be seen as follows: According to Equations \eqref{eq:majorityruledecodingsymmetric} and \eqref{eq:microentropy}, the macroscopic Shannon entropy of a symmetric MLB is symmetric with respect to $\macroprobability=1/2$. Therefore, the energetics of the erasure process from the initial state $\macroprobability_i=1/2$ to the final state $\macroprobability_f$ is the same as that of the erasure process from $\macroprobability_i=1/2$ to the final state $1-\macroprobability_f$. Hence, we will restrict to the symmetric majority-logic decoding characterized by Equation \eqref{eq:majorityruledecodingsymmetric} and to the erasure branch, $\macroprobability \leq 1/2$, in the following. To ease notation, we omit the explicit notation $\threshold=(\dimension+1)/2$ and $\macroprobability_{i} = \microprobability_{i}=1/2$.

As already pointed out earlier, the macroscopic entropy associated with an MLB can be thought of as a coarse-graining of the sum of the physical entropy of each microscopic unit contained in the MLB. We write the difference between the physical entropy of all microscopic units contained in the bit and the logical (Shannon-like) entropy of the macroscopic bit as 
\begin{align}
\mathbb{S}(\macroprobability,\dimension) = 
\dimension \microentropy \Big( \betafunction^{-1}_{\macroprobability} \Big( \tfrac{\dimension+1}{2},\tfrac{\dimension+1}{2} \Big) \Big) - \macroentropy(\macroprobability) .
\label{eq:entropydifference}
\end{align}
In Appendix \ref{sec:entropymontonicityproof}, a proof for the monotonic behavior of $\mathbb{S}(\macroprobability,\dimension)$ with respect to $\macroprobability$ in the case of symmetric majority-logic decoding is provided. In the range of $0 \leq \macroprobability \leq 1/2$, the entropy function $\mathbb{S}(\macroprobability,\dimension)$ is monotonically increasing, else it is decreasing monotonically. Evidently, $\mathbb{S}(\macroprobability,\dimension)$ takes its maximum value, $(\dimension-1)\ln 2$ if $\microprobability=\macroprobability=1/2$ and its minimum value, $ \mathbb{S}(\macroprobability,\dimension)=0 $, if $\microprobability=\macroprobability=0,1$. Hence, for symmetric majority-logic decoding the logical information underestimates or is equal to the physical one $\mathbb{S}(\macroprobability,\dimension) \geq 0 \; \forall \dimension$. As can be seen in Figure \ref{fig:heatandefficiencyreversible}a, such an inequality does not hold for the more general asymmetric case where $\threshold \neq (\dimension+1)/2$. In fact, the inequality $\mathbb{S}(\macroprobability,\dimension) \geq 0$ is physically equivalent to the well-known universal relationship between fully microscopic and coarse-grained Shannon entropies \cite{Esposito2012PRE}.
It is furthermore important to note that no general statement can be made about the changes of entropy for a given erasure process, since, according to Equation \eqref{eq:majorityruledecodingsymmetric} the final distribution for the microscopic unit is, in general, different from the one of the MLB.

From Equation \eqref{eq:macroefficiency} and the monotonicity of $\mathbb{S}(\macroprobability,\dimension)$ follows that the reversible erasure efficiency of the symmetric MLB is bounded as follows
\begin{align}
0 < \macroefficiency^{rev}(1/2,\macroprobability_f) \leq 1 ,  \label{eq:macroefficiencyreversible}
\end{align}
where the equality only holds when no information is erased, $\change \macroentropy(1/2,\macroprobability_f=1/2)=0 $. Hence we find that a SUB is always more efficient in reversible information erasure than a symmetric MLB.
This can be attributed to the majority-logic decoding that neglects some microscopic degrees of freedom and thus associates less logical entropy to the MLB than the cumulated physical entropy of the microscopic units that constitute the MLB.

We now state two limiting results for the reversible erasure efficiency of an MLB.
First, in case of perfect erasure, $\macroprobability_f=0$, we derive from Equations \eqref{eq:majorityruledecodingsymmetric} and \eqref{eq:macroefficiency} that the erasure efficiency simplifies to
\begin{align}
\macroefficiency^{rev}(1/2,0) = \frac{1}{\dimension} . \label{eq:macroefficiencyreversiblecompleteerasure}
\end{align}
Secondly, if the amount of erased information is small, $\macroprobability_f \approx \macroprobability_i = 1/2$, and $\dimension$ is large, the erasure~efficiency
\begin{align}
\macroefficiency^{rev}(1/2,\macroprobability_f) = \frac{2}{\pi} + \mathcal{O} \bigg( \left| \frac{1}{2}-\macroprobability_f \right|^{2} \bigg) ,
\label{eq:macroefficiencyreversiblesmallerasure}
\end{align}
becomes independent of the final macroscopic probability $\macroprobability_f \approx 1/2 $ as can be seen in Figure \ref{fig:heatandefficiencyreversible}b.

\section{Symmetric Majority-Logic Decoding under Finite-Time Erasure Protocol}
\label{sec:finitetimeprocesses}

Since we have captured the phenomenology of the reversible majority-logic decoding in the previous section, we now proceed by studying more realistic, finite-time information erasure processes. In dynamical processes the heat generation depends on the erasure protocol and the specific model for the microscopic units. In this section, we formulate finite-time erasure processes for two commonly employed microscopic models: We consider a two-state system with either Arrhenius rates or Fermi rates and denote in both cases by $\microprobability(t)$ the probability of the unit to be in state $1$ at time $t$.

\subsection{Preliminaries}
\unskip
\subsubsection{Master Equation}
The transition rate from state $0$ to state $1$ and vice versa is referred to as $\microrates_{10}(t)$ and $\microrates_{01}(t)$, respectively, which depends, in general, on time via the erasure protocol. We assume that the process is Markovian, such that the dynamics of $\microprobability(t)$ is ruled by a master equation:
\begin{align}
\dot{\microprobability}(t) = [1-\microprobability(t)] \, \microrates_{10}(t) - \microprobability(t) \, \microrates_{01}(t) , \label{eq:masterequation}
\end{align}
with the transition rates satisfying the local detailed balance relation
\begin{align}
\frac{\microrates_{10}(t)}{\microrates_{01}(t)}=\euler^{ - \invtemperature \, \Delta \epsilon(t) },\label{eq:localdetailedbalance}
\end{align}
where $\Delta \epsilon(t)$ is the energy gap from state $0$ to $1$ that is modulated in time according to the specific erasure protocol.
With Equation \eqref{eq:localdetailedbalance} the master Equation \eqref{eq:masterequation} can be cast into the form
\begin{align}
\dot{p}(t)=\left[\euler^{- \invtemperature \, \Delta \epsilon(t) } -(1 + \euler^{-\invtemperature\, \Delta \epsilon(t) })\microprobability(t)\right]\microrates_{01}(t) . 
\label{eq:masterequationrecast}
\end{align}

\subsubsection{Finite-Time Erasing}
No protocol can achieve a perfect erasure corresponding to $\microentropy(\microprobability_f)=\microentropy(\macroprobability_f)=0$, since this requires an infinite amount of time \cite{Giovanni2013PRE}. This can be seen as follows:
For a given microscopic model, there are several ways to decrease the microscopic probability $\microprobability(t)$ at different speeds, which corresponds to a protocol that ensues different amounts of heat and thus a different erasure efficiency.
This, in turn, implies that the time required to transform the initial probability $\microprobability_i(0) = 1/2$ into the final one $\microprobability_f(\tau) \equiv \microprobability_f $ is cannot be smaller than a minimal time $\tau_c$
\begin{align}
\tau \geq \tau_c = - \log 2 \microprobability_f , \quad  \microprobability_f < \frac{1}{2} ,
\label{eq:criticaltime}
\end{align}
As $\microprobability_f \to 0$, the minimal time diverges, $\tau_c \to \infty $, such that perfect erasure of a finite initial amount of information can only be realized by infinite-time protocols.

Equation \eqref{eq:criticaltime} can be rearranged to obtain a lower bound for the final probability obtained after an erasure with fixed duration $\tau$ as follows
\begin{align}
\microprobability_f \geq \microprobability_c = \frac{1}{2} \euler^{-\tau_c}, \quad \microprobability_f < \frac{1}{2},
\label{eq:criticalmicroprobability}
\end{align}
Consequently, in finite-time information erasure processes, the Shannon entropy of the final state $\microentropy(\microprobability_f)$ can be seen as the erasure error. According to Equation \eqref{eq:criticalmicroprobability}, the minimal erasure error of the process is $\microentropy(\microprobability_c)$ and the erasable information within $\tau_c$ is $\change \microentropy(\microprobability_f,1/2) = \log 2 - \microentropy(\microprobability_{c})$. 
According to Equations~\eqref{eq:criticaltime} and \eqref{eq:criticalmicroprobability}, the bounds $\tau_c$ and $\microprobability_{c}$ are rate- and thus model-independent.

These results hold at the level of a microscopic unit. We now proceed by discussing the bounds on the level of macroscopic bits. First, since each SUB consists of only one microscopic unit, Equations~\eqref{eq:criticaltime} and \eqref{eq:criticalmicroprobability} are also applicable to the macroscopic quantities, that is 
\begin{align}
\tau \geq \tau_c^s = \tau_c = - \log 2 \macroprobability_f , \quad  \macroprobability_f \geq \macroprobability_c^{\single} = \microprobability_c = \frac{1}{2} \euler^{-\tau}, \quad \macroprobability_f < \frac{1}{2} .
\label{eq:criticalmacroprobability} 
\end{align}
To calculate the lower bound on the macroscopic erasure time $\tau_{c}^{\majority}(1/2,\macroprobability_{f},\dimension)$ for a symmetric MLB, we recall that a macroscopic erasure from $\macroprobability_{i}=1/2$ to $\macroprobability_{f}$ is achieved by the corresponding erasure from $\microprobability_{i} =1/2$ to $\microprobability_{f}$ in each microscopic unit contained in the MLB.
Therefore, one has for the minimal erasure time in an MLB
\begin{align}
\tau_{c}^{\majority}(1/2,\macroprobability_{f},\dimension)=\tau_{c}(1/2,\microprobability_{f}),\label{eq:criticaltimemacroscopic}
\end{align}
where $\microprobability_f$ and $\macroprobability_f$ are related to each other via the symmetric majority-logic decoding in Equation~\eqref{eq:majorityruledecodingsymmetric}.
Hence, to compute $\tau_c^{\majority}$, the final microscopic probability $\microprobability_f$ needs to be determined via inversion of Equation \eqref{eq:majorityruledecodingsymmetric} and plugged into Equation \eqref{eq:criticaltime}. Next, using Equations \eqref{eq:majorityruledecodingsymmetric} and \eqref{eq:criticalmicroprobability} , one straightforwardly obtains the minimal final probability $\macroprobability_{c}^{\majority}(\tau,\dimension)$ after erasure time $\tau$. 
From Equation~\eqref{eq:majorityruledecodingsymmetric} follows the inequalities
\begin{align}
\tau_{c}^{\majority}(\macroprobability_{f},\dimension) \leq \tau_{c}^{\single}(\macroprobability_{f}), \quad \macroprobability_{c}^{\majority}(\tau,\dimension) \leq \macroprobability^{\single}_{c}(\tau),
\label{eq:criticalmicrovsmacro}
\end{align}
which suggest that majority-logic decoding can both accelerate the erasing process and additionally reduce the minimal erasure error.
Finally, for finite-time erasure processes we define the erasure power for a SUB and MLB as follows
\begin{align}
\micropower =  \frac{\change \microentropy(1/2, \microprobability_f)}{\tau}, \quad
\macropower =  \frac{\change \macroentropy(1/2, \macroprobability_f)}{\tau} ,\label{eq:power}
\end{align}
respectively.

\subsubsection{Arrhenius-Rates Unit Model}

The Arrhenius model consists of two potential well, which are regarded as states 0 and 1, respectively, that are separated by a barrier. The energy gap between state 0 and 1 is defined as $\Delta \epsilon$.
To be consistent with the assumption of Arrhenius transition rates, the energy barrier height of the potential well associated with state 1 is assumed constant, $\epsilon$, throughout the erasure process. This setup is different from the one used in Ref.~\cite{Jun2014}, where the two states are merged together and then separated during the erasure.
Here, the transition rates read
\begin{align}
\microrates^A_{01}(t)=r_{0} \, \euler^{- \invtemperature \epsilon }, \quad \mathrm{ and } \quad \microrates_{10}^A(t)=r_{0} \, \euler^{- \invtemperature [ \epsilon + \Delta \epsilon(t) ] },
\label{eq:arrheniusrates}
\end{align}
where $r_{0}$ is a constant setting the time scale of the process. Then, the master Equation \eqref{eq:masterequationrecast} can be written as follows
\begin{align}
\dot{\microprobability}(t) = \euler^{- \invtemperature \Delta \epsilon(t) } - \big[ 1+\euler^{-\invtemperature \Delta \epsilon(t) } \big] \, \microprobability(t) ,
\label{eq:masterequationarrhenius}
\end{align}
where the constant $r_{0} \, \euler^{- \invtemperature \epsilon}$ is absorbed into the time scale.

\subsubsection{Fermi-Rates Unit Model}

The Fermi-rates unit model can be experimentally realized via a single quantum dot with a single energy level $\energy$ that is in contact with a moving metallic lead corresponding to a time-dependent chemical potential $\chempot(t)$ and a heat bath at inverse temperature $\invtemperature$ \cite{Esposito2011EPL}. If the dot is filled by an electron we consider the unit to be in state $1$, else $0$. The transition rate for an electron leaving or entering the dot reads
\begin{align}
\microrates_{01}^F(t)= \frac{r_{0}}{\euler^{- \invtemperature \, \change \epsilon(t) } +1 }, \quad
\microrates_{10}^F(t)= \frac{r_{0}}{\euler^{ \invtemperature \, \change \epsilon(t) } +1 } ,\label{eq:fermirates}
\end{align}
respectively, where $\Delta \epsilon(t) \equiv \energy-\chempot(t)$ represents the energy barrier to enter the dot and $r_0$ constant setting the time scale of the process. Then, the master Equation \eqref{eq:masterequationrecast} can be written as follows
\begin{align}
\dot{\microprobability}(t) = - \microprobability(t) + \frac{1}{\euler^{\invtemperature \, \change \epsilon(t)} +1 }  , 
\label{eq:masterequationfermi}
\end{align}
where the constant $r_{0}$ is absorbed into the time scale. 
Since the second term in Equation \eqref{eq:masterequationfermi} is bounded between 0 and 1, the fastest way to decrease the microscopic probability $\microprobability(t)$ is realized by $\dot{\microprobability}(t) = - \microprobability(t)$, which corresponds to a protocol that ensues a divergent heat generation and thus a vanishing erasure efficiency.

\subsection{Constant State-Energy Protocol}
\unskip
\subsubsection{Variable Erasure Duration}
\begin{figure}[h!]
\begin{center}

\begin{tabular}{c}
\includegraphics[scale=0.95]{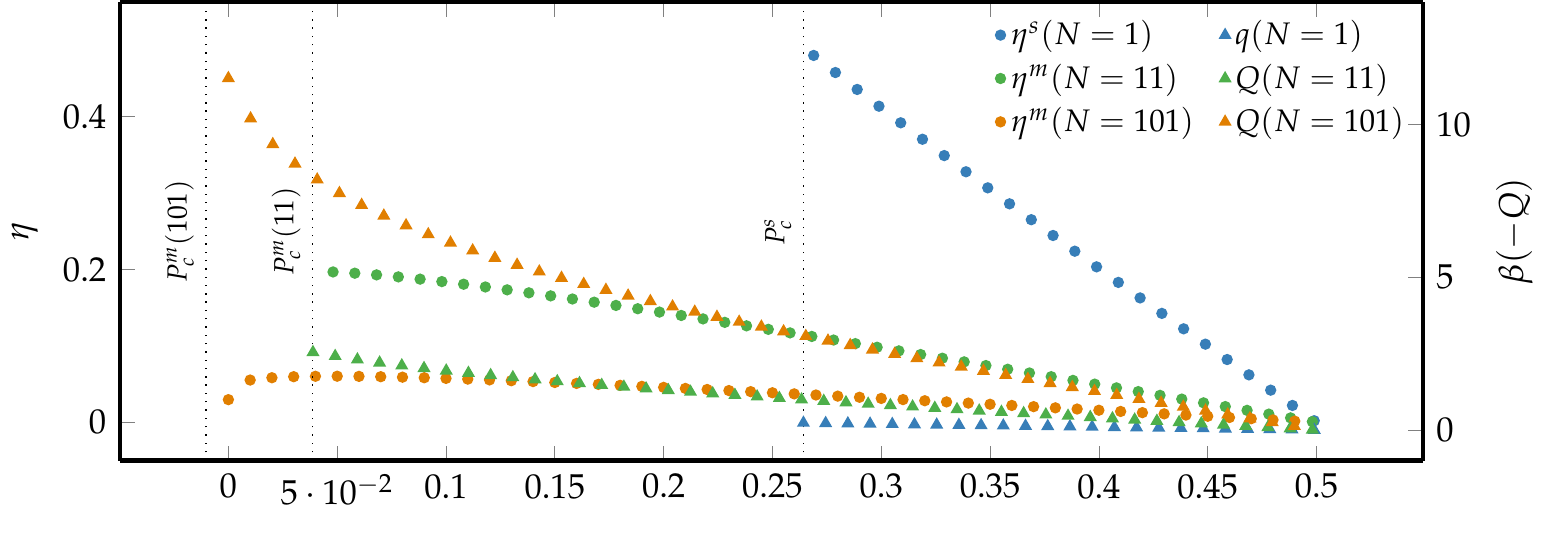}\\
(\textbf{a})\\
\includegraphics[scale=0.95]{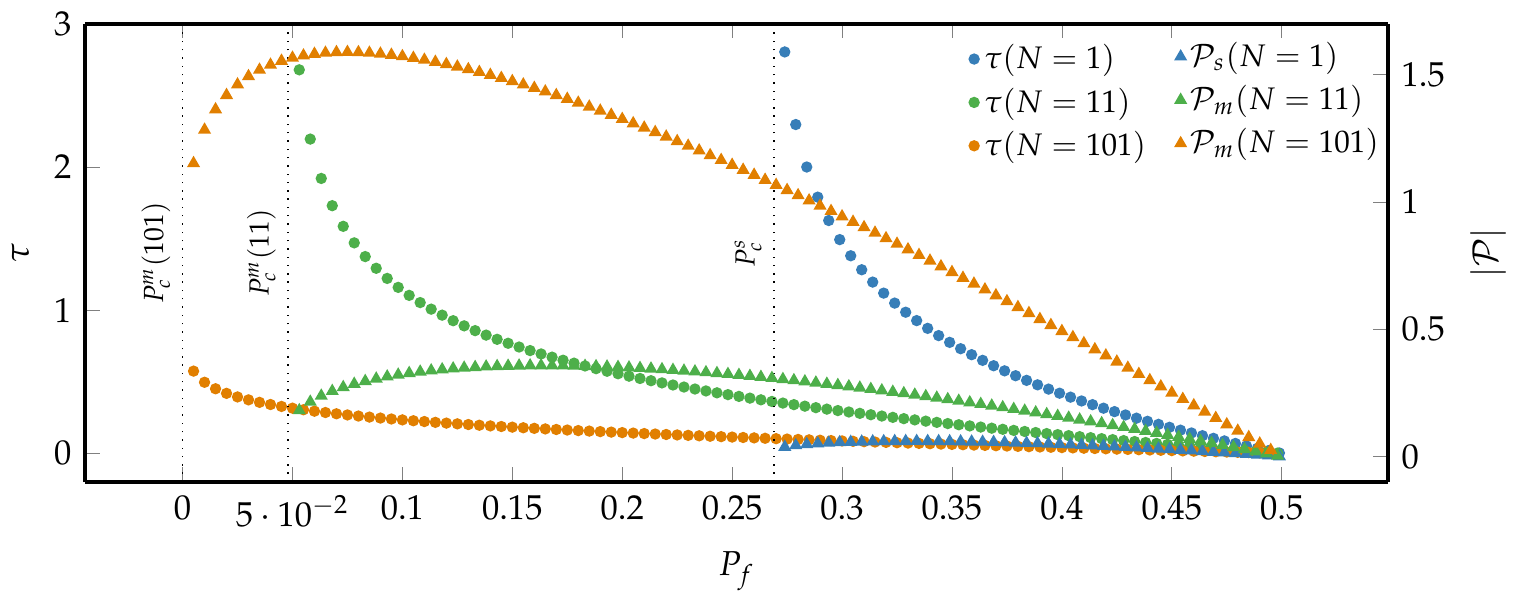}\\
(\textbf{b})\\
\end{tabular}

\caption{(\textbf{a}) Comparison of both the modulus of heat dissipation, $- \invtemperature \macroheat$, and the associated erasure efficiency between a SUB and a symmetric MLB for different $\dimension$. (\textbf{b}) Comparison of both the erasure time and the modulus of the power, $|\power|$, between a SUB and a symmetric MLB for different $\dimension$. The data was generated by using the generic solution of the master Equation \eqref{eq:masterequationsolutionfixedenergy} and setting $\microrates_{01}, \invtemperature \, \epsilon \equiv 1$. The dotted vertical lines correspond to the minimal erasure error $\macroprobability_c^{s}$ and $\macroprobability_c^{m}(\dimension)$ for the systems under~consideration. \label{fig:finitetimeconstantenergy}  }
\end{center}
\end{figure}

With the tools to address finite-time information processing at hand, we want to start with the simplest erasure protocol given by an instantaneous switching of the energy gap to the same value $\Delta \epsilon$ at time $t=0$.
For generic transition rates with constant energy gaps, $\microrates_{01}(\change \epsilon)$, the master equation of a microscopic unit from Equation \eqref{eq:masterequationrecast} is solved by
\begin{align}
\microprobability(t)=\frac{1}{2} \tanh \left( \frac{\invtemperature \Delta \epsilon}{2} \right) \euler^{-\left( 1 + \euler^{- \invtemperature \Delta \epsilon } \right) \microrates_{01}(\Delta\epsilon) \, t } + \frac{1}{1 + \euler^{\invtemperature \Delta \epsilon }} ,
\label{eq:masterequationsolutionfixedenergy}
\end{align}
where we used the initial condition $\microprobability(0)=1/2$.
In the infinite-time limit the probability converges to the lower bound
\begin{align}
\microprobability(\infty) = \frac{1}{1 + \euler^{\invtemperature \Delta \epsilon }} .
\label{eq:masterequationsolutionfixedenergyinfinitetime}
\end{align}
We emphasize that for fixed $\Delta \epsilon$ the definitions \eqref{eq:criticaltime} and \eqref{eq:criticalmicroprobability} no longer apply. However, Equation~ \eqref{eq:masterequationsolutionfixedenergyinfinitetime} represents also a bound on the minimal erasure error and thus plays a similar role as $\microprobability_{c}$. From Equation~\eqref{eq:masterequationsolutionfixedenergy} we obtain the following expression for the erasure duration 
\begin{align}
\tau = - \frac{1}{(1 + \euler^{- \invtemperature \Delta \epsilon }) \, \microrates_{01}(\Delta\epsilon)}  \ln{ \left( 2 \frac{(1 + \euler^{\invtemperature \Delta \epsilon }) \microprobability_{f} - 1}{ \euler^{\invtemperature \Delta \epsilon } - 1  } \right) }.\label{eq:criticaltimefixedenergy}
\end{align}
The heat dissipated by the microscopic unit reads
\begin{align}
\microheat =  \int_{0}^{\tau} \Delta \epsilon \dot{\microprobability}(t) \, \d t = \Delta \epsilon \int_{\microprobability_i}^{\microprobability_f} \! \d \microprobability = \bigg( \microprobability_{f} - \frac{1}{2} \bigg) \Delta\epsilon , \label{eq:microheatfixedenergy}
\end{align}
which, with Equation \eqref{eq:macroheat}, results in the total heat generated by the SUB and MLB
\begin{align}
\microheat &= \bigg( \microprobability_{f} - \frac{1}{2} \bigg) \Delta \epsilon, \quad
\macroheat = \dimension  \left[ \betafunction^{-1}_{\macroprobability_{f}} \bigg( \frac{N+1}{2},\frac{N+1}{2} \bigg) - \frac{1}{2} \right] \Delta \epsilon , 
\label{eq:macroheatfixedenergy}
\end{align}
respectively. With the definitions in Equations \eqref{eq:microefficiency}, \eqref{eq:macroefficiency} and \eqref{eq:power} one has for the macroscopic erasure~power
\begin{align} \label{eq:powerfixedenergy}
\power^{\single} \bigg( \frac{1}{2},\microprobability_f \bigg) = \frac{ \change \microentropy(1/2,\microprobability_f) }{ \tau }, \quad
\power^{\majority} \bigg( \frac{1}{2},\macroprobability_f \bigg) = \frac{ \change \macroentropy(1/2,\macroprobability_f) }{ \tau } ,
\end{align}
and the macroscopic efficiencies 
\begin{align} \label{eq:efficiencyfixedenergy}
\microefficiency \bigg( \frac{1}{2},\microprobability_f \bigg) = \frac{ \change \microentropy(1/2,\microprobability_f) }{ \invtemperature \, \microheat }, \quad
\macroefficiency \bigg( \frac{1}{2},\macroprobability_f \bigg) = \frac{ \change  \macroentropy(1/2,\macroprobability_f ) }{ \invtemperature \, \macroheat }  .
\end{align}
The results of the finite-time erasure process for a fixed energy gap are depicted in Figure \ref{fig:finitetimeconstantenergy}.

 As can be seen in panel (a), to perform an erasure with the same erasure error the SUB dissipates less heat and thus has a higher efficiency than the symmetric MLBs, for which the heat production increases and the efficiency decreases with growing $\dimension$.
Hence the SUB is more efficient than the symmetric majority-logic decoding, as already observed in the reversible case, cf. Figure \ref{fig:heatandefficiencyreversible}.
However, Figure \ref{fig:finitetimeconstantenergy}b shows that the erasure duration is reduced, and the erasure power thus enhanced by employing the majority-logic decoding for large ensembles of microscopic units.
The minimal erasure error characterized by Equation \eqref{eq:masterequationsolutionfixedenergyinfinitetime} is indicated by the dotted vertical lines that correspond to the minimal final probability $\macroprobability_c^{m}(\dimension)$.
Therefore, as already derived in Equation \eqref{eq:criticalmicrovsmacro}, majority-logic decoding reduces the minimal erasure error that goes to zero as $\dimension$ becomes large corresponding to perfect erasure.

\subsubsection{Fixed Erasure Duration}
We now assume that the explicit form of the transition rates of the microscopic unit model is known and compare the performance of the two types of macroscopic bits under the protocol with fixed erasing time and instantaneous switching of the energy gap. For a specific erasure process, the erasure time and thus the erasure power are equal for the two macroscopic bits, hence we restrict the discussion to the macroscopic erasure efficiencies.
The microscopic dynamics of a Fermi-rates unit~reads
\begin{align}
\microprobability(t) = \frac{1}{2} \tanh \left( \frac{\invtemperature \Delta \epsilon}{2} \right) \euler^{-t} + \frac{1}{1 + \euler^{\invtemperature \Delta \epsilon }} .
\label{eq:masterequationsolutionfixedenergyfixedtimefermi}
\end{align}
Plugging the erasure time $\tau$ into Equation \eqref{eq:masterequationsolutionfixedenergyfixedtimefermi}, yields the final microscopic probability and the erasure error. The calculation of the generated heat, erasure power and efficiency both on the microscopic and macroscopic level is analogous to the one in the last section. As the specific dynamics of the unit is known and $\tau$ is fixed, the definition of minimal final probability in Equation \eqref{eq:criticalmicroprobability} is valid under this~protocol.

\begin{figure}[h!]
\begin{center}

%
%
%
%
%
%
%

\includegraphics[scale=1]{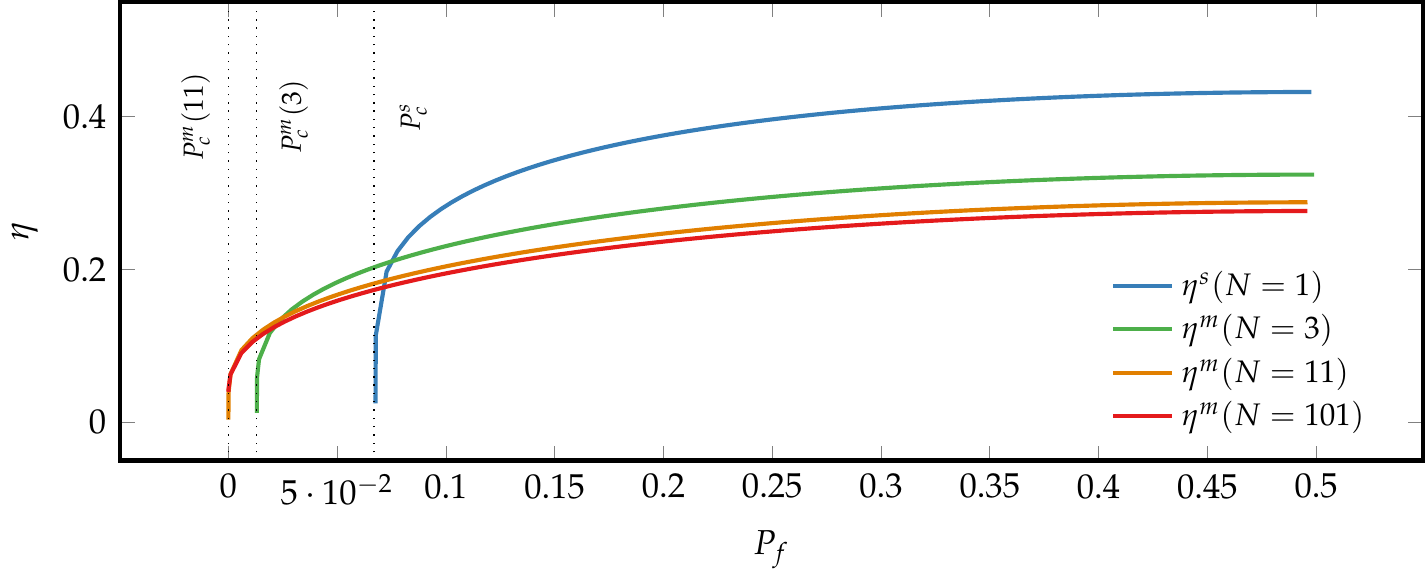}

\caption{Comparison between the erasure efficiency of a SUB and symmetric MLBs for different $\dimension$ and instantaneously switched $\Delta \epsilon$ using the Fermi-rates unit model and fixed erasure time $\tau =2$. The vertical dotted lines correspond to the minimal final macroscopic probabilities $\macroprobability_c^{s}$ and $\macroprobability_c^{m}(\dimension)$, where for $\dimension = 11,101$ the probabilities are too close to zero to be distinguished. \label{fig:finitetimeconstantenergyfixedtime}}
\end{center}
\end{figure}

Figure \ref{fig:finitetimeconstantenergyfixedtime} shows that the symmetric majority-logic decoding has additional advantages for finite-time erasing:
As already observed earlier, the minimal erasure error of the symmetric MLB is smaller than that of the SUB and approaches $0$ with increasing $\dimension$ as illustrated by the vertical dotted lines corresponding to the minimal final probability $\macroprobability_c^{m}(\dimension)$ after the erasure time $\tau$. More importantly, the symmetric MLBs are more efficient in the region of small-erasure error region, $\macroprobability_f \approx \macroprobability_c^{s}$, as opposed to the region of large erasure error where the SUB is more efficient. 
We find that for Arrhenius rates the results are qualitatively similar and thus omitted.

To sum up, Figures \ref{fig:finitetimeconstantenergy} and \ref{fig:finitetimeconstantenergyfixedtime} illustrate the important result that the precision-speed-efficiency trade-off in finite-time information erasure processes is lifted by symmetric majority-logic decoding.

\subsection{Optimal Erasure Protocol}
\label{sec:optimalprotocol}

In view of applications, the least work-intense erasure processes are of particular interest. The general method to determine the optimal erasure protocol that minimizes the generated heat in a Fermi-rates unit model has been established in Ref. \cite{Esposito2010EPL}. The detailed derivation of the heat-minimizing protocol for Arrhenius rates is deferred to Appendix \ref{sec:optimalprotocolarrheniusderivation}. Given the optimal protocol, both the microscopic and macroscopic heat dissipation, erasure power and efficiency follow readily from Equations \eqref{eq:microheatfixedenergy}, \eqref{eq:powerfixedenergy} and \eqref{eq:efficiencyfixedenergy}.

\begin{figure}[h!]
\begin{center}

\includegraphics[scale=0.99]{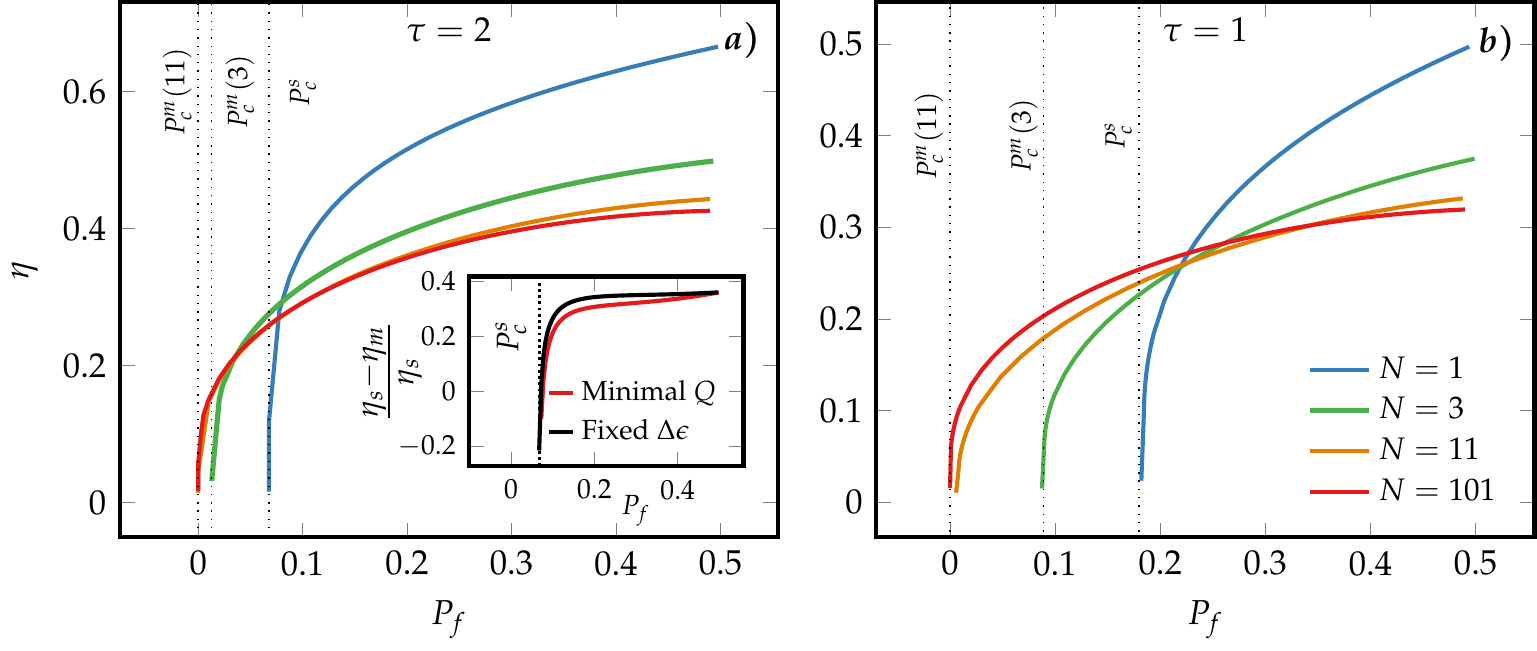}

\caption{Comparison between the erasure efficiency of a SUB and of a symmetric MLB based on the Arrhenius model for different $\dimension$ under the optimal erasure protocol with the erasure time $\tau=2$ [panel~(\textbf{a})] and $\tau=1$ [panel (\textbf{b})]. The vertical dotted lines correspond to the minimal final macroscopic $\macroprobability_c^{s}$ and probabilities $\macroprobability_c^{m}(\dimension)$, where for $\dimension = 11,101$ the probabilities are too close to zero to be distinguished.
The inset in panel (\textbf{a}) shows the erasure efficiency of a SUB compared to a symmetric MLB one, $(\microefficiency-\macroefficiency)/\microefficiency$ for $\dimension=101$ using the optimal and the fixed energy protocol.
\label{fig:optimalprocotolarrheniusefficiencies} } 

\end{center}
\end{figure}

The erasure efficiency of a SUB and a symmetric MLBs for different $\dimension$ are compared in Figure~ \ref{fig:optimalprocotolarrheniusefficiencies}. It is important to note that the advantages  in terms of information erasure inherent to symmetric majority-logic decoding discussed in the previous section are preserved and enhanced by the optimal erasure protocol:
First, the minimal erasure error is strongly reduced for a symmetric MLB at the expense of an erasure efficiency that decreases with increasing $\dimension$ in the regime of small erasure, $\macroprobability_f \gg \macroprobability_c^{s} $. Conversely, for small-erasure error, $\macroprobability_f \approx \macroprobability_c^{s} $, this relation between the macroscopic efficiencies is inverted. The inset in panel (a) that depicts the relative erasure efficiency between a SUB and a symmetric MLB, $ ( \microefficiency - \macroefficiency )/\microefficiency $, reveals that the range of small-erasure error probabilities over which this holds true is increased for the optimal protocol compared to the fixed energy protocol.
Comparing furthermore panels (a) and (b) also shows that this range of small erasure-error probabilities is increasing with decreasing erasure time $\tau$ and increasing $\dimension$.

In the limit of low and high dissipation, analytical results can be obtained for the optimal protocol, where we will focus on the Arrhenius-rates unit model.
The high-dissipation limit corresponds to an erasure duration $\tau$ that approaches the minimal erasure time $\tau_c$. Thus, the parameter $K$ in Equation~\eqref{eq:lagrangeequation}, which represents the degree of irreversibility, is diverging in this limit. Using Equations~\eqref{eq:erasuretimeexpression} and \eqref{eq:erasuretimeexpressionfunctiondefinition}, the parameter $K$ can be expressed as
\begin{align}
K^{high}(\microprobability_{f},\tau) = \frac{1}{2} \frac{1 - 2 \microprobability_{f}}{\tau+\ln{(2\microprobability_{f})}} .
\label{eq:parameterhighdissipation}
\end{align}
Plugging Equation \eqref{eq:parameterhighdissipation} into Equation \eqref{eq:heatfunctionalsolution},
one has for the heat dissipation
\begin{align}
\invtemperature \, \microheat^{high}(\microprobability_{f},\tau) = \left( \frac{1}{2} - \microprobability_{f} \right) \ln \left[ \frac{1}{2} \frac{1 - 2 \microprobability_{f}}{\tau + \ln{(2\microprobability^{f})}} \right]. \label{eq:heathighdissipation}
\end{align}
We verify that $K, \microheat \to \infty$ as $\tau \to \tau_c = - \ln (2 \microprobability_f)$ and that Equation \eqref{eq:heathighdissipation} also represents the solution for a Fermi-rates unit.
In the low-dissipation limit, the erasure duration $\tau$ is large compared to $\tau_{c}$, hence the parameter $K$ is small. Using Equations \eqref{eq:erasuretimeexpression} and \eqref{eq:erasuretimeexpressionfunctiondefinition}, the parameter $K$ can be expressed as
\begin{align}
K^{low}(\microprobability_{f},\tau) = 8 \left[ \frac{1 - \sqrt{2 \microprobability_{f}} }{2\tau + \ln (2\microprobability_{f} ) } \right]^2 . \label{eq:parameterlowdisspation}
\end{align}
Plugging Equation \eqref{eq:parameterlowdisspation} into Equation \eqref{eq:heatfunctionalsolution},
one has for the heat dissipation
\begin{align}
\invtemperature \, \microheat^{low}(\microprobability_{f},\tau) = \frac{2 \left(1 - \sqrt{2\microprobability_{f}} \right)^{2}}{\tau} - \Delta \microentropy(1/2,\microprobability_f) . \label{eq:heatlowdissipation}
\end{align}
Significantly, the first term in Equation \eqref{eq:heatlowdissipation}, which could be interpreted as the irreversible dissipation, is consistent with the low-dissipation assumption made in Refs. \cite{Aurell2012JSP,Schmiedl2008EPL,Esposito2010PRL}.
Here, the expression for the heat in Equation \eqref{eq:heatlowdissipation} differs from that of a Fermi-rates unit at low dissipation.
The heat dissipated by a SUB in the high ($\microheat^{high}$) and low-dissipation limit ($\microheat^{low}$) are the same as those of the microscopic unit and therefore given by Equations \eqref{eq:heathighdissipation} and \eqref{eq:heatlowdissipation}, respectively.
The heat dissipated by a symmetric MLB in the high ($\macroheat^{high}$) and low-dissipation limit ($\macroheat^{low}$) are readily derived using Equation \eqref{eq:majorityruledecodingsymmetric} with results from Equations \eqref{eq:heathighdissipation} and \eqref{eq:heatlowdissipation}.
For small-erasure ($\microprobability_{f} \to 1/2 $), the expression for the heat in the low-dissipation limit in Equation \eqref{eq:heatlowdissipation} simplifies to
\begin{align}
\microheat^{low} \approx 2 \left(1 + \frac{1}{\tau} \right) \left( \frac{1}{2} - \microprobability_{f} \right)^{2} . \label{eq:heatlowdissipationsmallerasure}
\end{align}
Using the approximation for the majority-logic decoding in the limit of small-erasure and large-$\dimension$ in Equation \eqref{eq:micromacroprobabilityapproximation}, the ratio $\kappa$ between the erasure efficiency of a symmetric MLB and a SUB
\begin{align}
\kappa \equiv \frac{\macroefficiency(\macroprobability_{f},\dimension,\tau)}{\microefficiency(\microprobability_{f},\tau)} \approx \frac{2}{\pi}, \label{eq:efficiencyrationlowdissipation}
\end{align}
is independent of the erasure duration $\tau$ and final erasure error. 

\begin{figure}[h!]
\begin{center}

\includegraphics[scale=0.99]{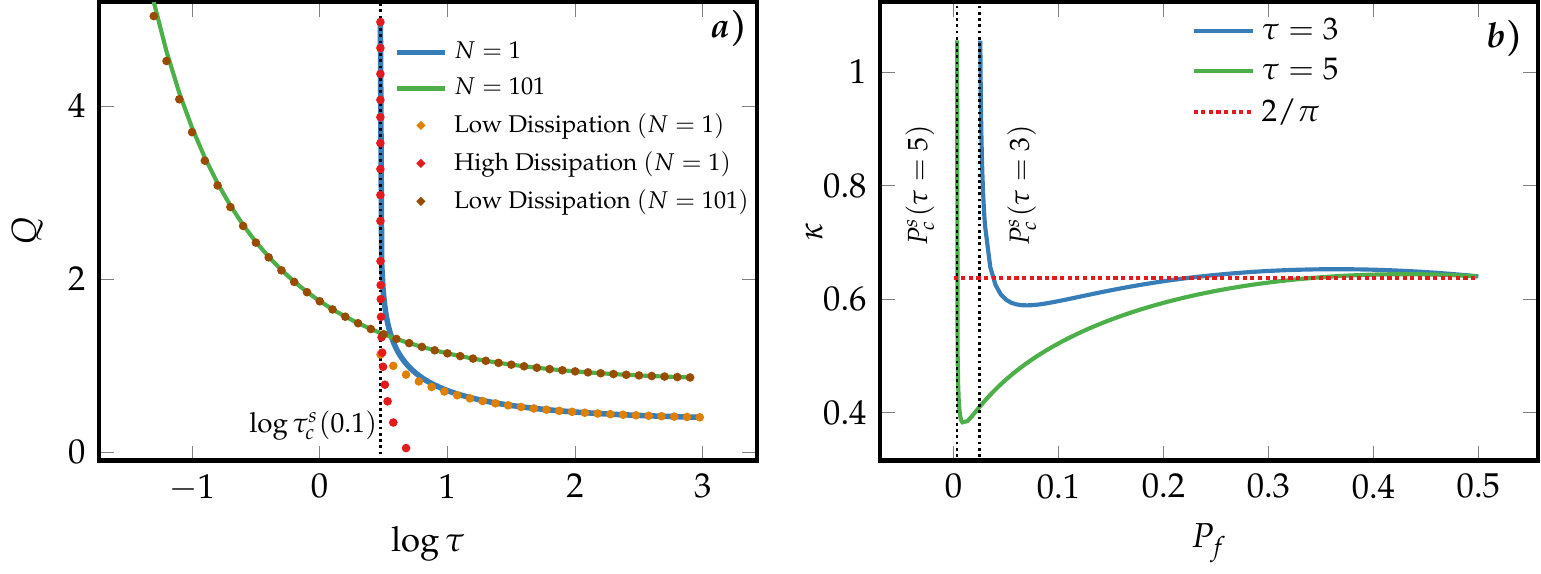}

\caption{(\textbf{a}) Comparison between the dissipated heat of a SUB and a symmetric MLB ($\dimension = 101$) based on Arrhenius-rates units as a function of erasure duration $\log \tau$ for the optimal erasure protocol with $\macroprobability_f =0.1$. In addition to the full numerical solution of Equation \eqref{eq:heatfunctionalsolution}, the analytic low- and high-dissipation solutions in Equations \eqref{eq:heathighdissipation} and \eqref{eq:heatlowdissipation} are displayed. The vertical dotted line corresponds to the logarithm of the critical time $\log \tau_c^s(0.1)$ of a single microscopic unit for this specific erasure process. 
(\textbf{b}) Ratio between the erasure efficiencies of a symmetric MLB ($\dimension=101$) and a SUB as a function of the final macroscopic probability $\macroprobability_f$ for the optimal protocol with the erasure times $\tau = 3,5$. The red dotted line corresponds to the small-erasure and large-$\dimension$ approximation given by Equation~ \eqref{eq:efficiencyrationlowdissipation} and the vertical black dotted lines represent the minimal erasure error $\macroprobability_c^{s}(\tau)$.
\label{fig:optimalprocotolarrheniusheat}  } 
\end{center}
\end{figure}

Figure \ref{fig:optimalprocotolarrheniusheat}a compares the dissipated heat of a SUB and a symmetric MLB for different erasure durations $\log \tau$ using the optimal protocol applied to a given erasure process from $\macroprobability_i = 1/2 $ to $\macroprobability_f $.

Additionally, the approximate solutions for the low- and high-dissipation limit in Equations \eqref{eq:heathighdissipation} and \eqref{eq:heatlowdissipation} are overlaid.
As can be seen, the symmetric MLB under the optimal protocol is still more efficient in the fast-erasure region, where the erasing duration $\tau$ approaches the minimal erasure time $\tau_{c}^s(\microprobability_f)$ of a SUB. Except for extremely fast-erasure processes, the full numerical solution of the heat dissipated by the symmetric MLB is in excellent agreement with the low-dissipation approximation in Equation \eqref{eq:heatlowdissipationsmallerasure}. This suggests that the calculation of the dissipated heat of a symmetric MLB built upon microscopic Arrhenius-rate units under the optimal erasure protocol can be simplified by using the more convenient Equation \eqref{eq:heatlowdissipationsmallerasure} instead of the numerically more involving procedure elaborated in Appendix \ref{sec:optimalprotocolarrheniusderivation}. 

In Figure \ref{fig:optimalprocotolarrheniusheat}b the ratio between the efficiency of a symmetric MLB and a SUB are compared for the same erasure process. In agreement with the observation made in panel (a) that the heat dissipation in the fast-erasure region is significantly reduced for a symmetric MLB, we note in panel (b) that the erasure efficiency of a symmetric MLB is considerably higher than that of a SUB ($\kappa >1$) in this region. It is interesting to notice that even though the approximate expression for the efficiency ratio in Equation \eqref{eq:efficiencyrationlowdissipation} is theoretically only valid in asymptotic limit $ \macroprobability_f \rightarrow 1/2$, the full numerical solutions are in good agreement with the approximate solution ($2/\pi$) for a large range of values for $\macroprobability_f$.

We therefore conclude that with the aid of symmetric majority-logic decoding, the energetically optimized erasure processes can be accelerated, be performed more precise at a lower cost. Hence, for large-$\dimension$ symmetric MLBs the speed-precision-efficiency trade-off is significantly lifted. We remark that the optimization procedure (Appendix \ref{sec:optimalprotocolarrheniusderivation}) of the Fermi-rates unit model was studied in detail in~ \cite{Esposito2010EPL}. The results of the performance of a symmetric MLB built upon Fermi-rates units are qualitatively similar to those for the Arrhenius-rates unit model discussed above and thus omitted. It should however be emphasized that the advantages of employing the symmetric majority-logic decoding for information processing are preserved under a change between these two different microscopic models.

\section{Conclusions}
\label{sec:conclusion}
In this work, we studied the performance of majority-logic decoding in reversible and finite-time information erasure processes. To this end, we introduced two macroscopic bits, the single-unit and the majority-logic bit, that contain one and $\dimension$ microscopic binary units, respectively. The physical information stored inside the $\dimension$ microscopic units is translated into logical information stored in the macroscopic bit via the majority-logic decoding scheme that can be mathematically formulated via an incomplete regularized beta function.

We found that for reversible erasure protocols the SUB is always more efficient in erasing a given amount of information than the symmetric majority-logic bit.
Conversely, for finite-time information erasure processes that are naturally characterized by the constraints of finite erasure time and erasure error, the majority-logic bit exhibits multiple advantages compared to a single unit:
Both the erasure time can be reduced, and the erasure error can be compressed, hence accelerating and refining the erasure.
Remarkably, in the region of small erasure or fast erasure, the majority-logic unit also exhibits a higher efficiency. These advantages are preserved and enhanced as one changes from a simple (constant energy barrier) to the optimal protocol (minimal heat dissipation).
Hence, our results suggest that majority-logic decoding considerably lifts the speed-precision-efficiency trade-off in finite-time information erasure processes.
In view of applications, our work reveals a promising avenue to accelerate the information processing rather than focusing on the storage media which may become a severe limitation in the future semiconductor industry.
If one is restricted to a few SUBs per majority-logic bit, it would furthermore be interesting to consider a hierarchical system of majority-logic bits and to study if the thermodynamic benefits of majority-logic decoding can be further enhanced by such an iterative approach.

Finally, there is an important remark to be made, here. As discussed at the end of Section \ref{sec:coarsegraining} the majority-logic decoding can be thought of as a coarse-graining procedure. Consequently, the logical information stored in the macroscopic bit is not equal to the sum of microscopic physical information stored in the units contained in the bit. However, these hidden costs were not accounted for in the definitions of the macroscopic erasure power and macroscopic erasure efficiency. A more rigorous approach would possibly be to implement a physical mechanism that represents the majority-logic decoding, e.g., an interacting system of microscopic units that exhibits a phase transition. For instance, a two-dimensional Ising model or a majority-vote model of the kind found in Ref. \cite{tomte2005pre} could be used in the context of information erasure. The need to quantify the costs incurred by majority-logic decoding to validate the latter as a promising strategy to enhance information processing might stimulate future works.

\vspace{6pt}

\acknowledgments{We thank Christian Van den Broeck for suggesting investigating the majority-logic decoding and Zhanchun Tu for fruitful discussions and instructive suggestions. S.S. is grateful for the financial support of the China Postdoctoral Science Foundation (Grants NO. 2018M642124), the National Nature Science Foundation of China (Grants No. 11675017) and the Fundamental Research Funds for the Central Universities (Grants No. 2017EYT24) and China Scholarship Council.
T. H., G. D. and M. E. acknowledge funding by the National Research Fund, Luxembourg, in the frame of the AFR PhD Grant 2016, No. 11271777, in the frame of Project No. FNR/A11/02 and by the European Research Council project NanoThermo (ERC-2015-CoG Agreement No.~681456).
}

\appendix

\section{Equivalence of the Binomial Cumulative Distribution Function and \boldmath{$\betafunction$}}
\label{sec:proofbinomialbetaequality}

We start from the definition of the regularized incomplete beta function in \eqref{eq:betafunction} and solve the integral in the denominator via partial integration yielding
\begin{align}
\int_{0}^{1}t^{a-1}(1-t)^{b-1}\mathrm{d}t = \frac{(a-1)!(b-1)!}{(a+b-1)!} .
\end{align}
The integral in the numerator is determined via consecutive application of partial integration
{\small \begin{align}
\int_{0}^{x}t^{a-1}(1-t)^{b-1}\mathrm{d}t = \frac{1}{a} x^{a} (1-x)^{b-1} + \frac{b-1}{a(a+1)} x^{a+1} (1-x)^{b-2} + \ldots 
+ \frac{(b-1) (b-2) \cdot \ldots \cdot 2 \cdot 1}{a(a+1)\cdot \ldots \cdot (a+b-1)} x^{a+b-1}.
\end{align} }
Collecting results, one finds that 
\begin{align}
\betafunction_{x}(a,b) = \sum\limits_{k=a}^{a+b-1} {a+b-1 \choose a} x^{a} (1-x)^{b-1} .
\end{align}
If we set $a=\threshold,b=N-\threshold+1,x=p$, we immediately arrive at
\begin{align}
\macroprobability(\microprobability,N,\threshold)=\betafunction_{\microprobability}(\threshold,N+1-\threshold) ,
\end{align}
which proves Equation \eqref{eq:binomialbetaequality}.

\section{The Monotonicity of the Entropy Function \boldmath{$\mathbb{S}(\macroprobability,\dimension)$}}
\label{sec:entropymontonicityproof}
The first-order derivative of the function $\mathbb{S}(\microprobability,\dimension)$ introduced in Equation \eqref{eq:entropydifference} with respect to $\microprobability$ can be expressed as
\begin{align}
\frac{\partial \mathbb{S}(\microprobability,\dimension)}{\partial \microprobability} = \dimension \, \ln{\frac{1-\microprobability}{\microprobability}}-\frac{(1-\microprobability)^{\tfrac{\dimension-1}{2}}\microprobability^{\tfrac{\dimension-1}{2}}}{\mathcal{B} \left( \tfrac{\dimension+1}{2},\tfrac{\dimension+1}{2} \right)} \ln{\frac{1-f(\microprobability,\dimension)}{f(\microprobability,\dimension)}}  , 
\label{eq:entropydifferencefirstderivative}
\end{align}
where we write the regularized incomplete beta function with symmetric arguments as $f(\microprobability,\dimension) \equiv \betafunction_{\microprobability}\Big( \tfrac{\dimension+1}{2},\tfrac{\dimension+1}{2} \Big)$ and introduce the beta function $ \mathcal{B} \left( a,b \right) \equiv \int_{0}^{1}t^{a-1}(1-t)^{b-1} \, \mathrm{d}t $.
We find for the second-order derivative
\begin{equation}
\frac{\partial^{2} \mathbb{S} (\microprobability,\dimension)}{\partial \microprobability^{2}}=\frac{(\dimension-1)(1-\microprobability)^{\frac{\dimension-3}{2}}\microprobability^{\frac{\dimension-3}{2}}(2\microprobability-1) \ln{\frac{1-f(\microprobability,\dimension)}{f(\microprobability,\dimension)}} }{2\mathcal{B}(\frac{\dimension+1}{2},\frac{\dimension+1}{2})}
- \frac{\dimension}{\microprobability(1-\microprobability)}
\bigg[ 1 - 
\begin{pmatrix} \dimension \\ \tfrac{\dimension-1}{2} \end{pmatrix}^2
\frac{ \dimension (1-\microprobability)^{\dimension}\microprobability^{\dimension}}{(1-f)f} \bigg] .
\label{eq:entropydifferencesecondderivative}
\end{equation}
We easily verify that the first term on the right-hand side of Equation \eqref{eq:entropydifferencesecondderivative} is non-positive for any $\microprobability$.

To determine the sign of the second term on the right-hand side of Equation \eqref{eq:entropydifferencesecondderivative}, we first recast the term $(1-f)f$ into
\begin{align}
[1-f(\microprobability,\dimension)]f(\microprobability,\dimension) &= \sum_{i=\frac{\dimension+1}{2}}^{\dimension}\sum_{j=0}^{\frac{\dimension-1}{2}}{\binom {\dimension} i}{\binom {\dimension} j}\microprobability^{i+j}(1-\microprobability)^{2\dimension-(i+j)} \\
&=\sum_{m=\frac{\dimension+1}{2}}^{\dimension-1}\sum_{i=\frac{\dimension+1}{2}}^{m}{\binom {\dimension} i}{\binom {\dimension} {m-i}}[\microprobability^{m}(1-\microprobability)^{2\dimension-m}+\microprobability^{2\dimension-m}(1-\microprobability)^{m}]
 \; + \\ 
 &+ \sum_{i=\frac{\dimension+1}{2}}^{\dimension}{\binom {\dimension} {i}}{\binom {\dimension} {\dimension-i}}\microprobability^{\dimension}(1-\microprobability)^{\dimension},\label{eq:entropyproofone}
\end{align}
where we have introduced the index $m=i+j$. Using the inequality 
\begin{align}
\microprobability^{s}(1-\microprobability)^{4n+2-s}+\microprobability^{4n+2-s}(1-\microprobability)^{s} \geq 2\microprobability^{2n+1}(1-\microprobability)^{2n+1} ,
\end{align}
Equation \eqref{eq:entropyproofone} can be transformed into the inequality
\begin{align}
\left[1-f(\microprobability,\dimension)\right]f(\microprobability,\dimension)
\geq \left[ 2\sum_{m=\frac{\dimension+1}{2}}^{\dimension-1}\sum_{i=\frac{\dimension+1}{2}}^{m}{\binom {\dimension} i}{\binom {\dimension} {m-i}}+\sum_{i=\frac{\dimension+1}{2}}^{\dimension}{\binom {\dimension} {i}}{\binom {\dimension} {\dimension-i}}\right] \microprobability^{\dimension}(1-\microprobability)^{\dimension} ,
\label{eq:entropyprooftwo}
\end{align}
where the equal sign holds for $\microprobability=1/2$.

Furthermore, using the binomial theorem, we write
\begin{align}
2^{2\dimension-2}=&\left[\sum_{i=\frac{\dimension+1}{2}}^{\dimension}{\binom {\dimension} i}\right]\left[\sum_{j=0}^{\frac{\dimension-1}{2}}{\binom {\dimension} j}\right]\\
=&\sum_{m=\frac{\dimension+1}{2}}^{\dimension-1}\sum_{i=\frac{\dimension+1}{2}}^{m}{\binom {\dimension} i}{\binom {\dimension} {m-i}}+\sum_{i=\frac{\dimension+1}{2}}^{\dimension}{\binom {\dimension} i}{\binom {\dimension} {\dimension-i}}+\sum_{m=\dimension+1}^{\frac{3\dimension-1}{2}}\sum_{i=m-\frac{\dimension-1}{2}}^{\dimension}{\binom {\dimension} i}{\binom {\dimension} {m-\frac{\dimension-1}{2}}}\\
=&2\sum_{m=\frac{\dimension+1}{2}}^{\dimension-1}\sum_{i=\frac{\dimension+1}{2}}^{m}{\binom {\dimension} {m-i}}{\binom {\dimension} {i}}+\sum_{i=\frac{\dimension+1}{2}}^{\dimension}{\binom {\dimension} i}{\binom {\dimension} {\dimension-i}},
\end{align}
which is exactly equal to the prefactor in Equation \eqref{eq:entropyprooftwo}. Thus, we arrive at the inequality
\begin{align}
[1-f(\microprobability,\dimension)]f(\microprobability,\dimension)\geq 2^{2\dimension-2} \microprobability^{\dimension}(1-\microprobability)^{\dimension} .
\label{eq:entropyproofthree}
\end{align}
Substituting Equation \eqref{eq:entropyproofthree} into Equation \eqref{eq:entropydifferencesecondderivative}, we derive
\begin{align}
\frac{\partial^{2} \mathbb{S} (\microprobability,\dimension)}{\partial \microprobability^{2}}\leq 0 ,
\label{eq:entropydifferencecurvature}
\end{align}
by using the inequality
\begin{align}
\frac{2}{\pi}<\frac{4\dimension}{2^{2\dimension}}{\binom {\dimension-1} {(\dimension-1)/2}}^{2}\leq 1 .
\end{align}

Since $\partial \mathbb{S}(\microprobability,\dimension)/\partial \microprobability$ is a monotonically decreasing function with respect to $\microprobability$, it follows from $\partial \mathbb{S}(\microprobability,\dimension)/\partial \microprobability|_{\microprobability=1/2} = 0$ that $\partial \mathbb{S}(\microprobability,\dimension)/\partial \microprobability$ is positive (negative) for $\microprobability < 1/2$ ($\microprobability > 1/2$).
Therefore, $\mathbb{S}(\microprobability,\dimension)$ is monotonically increasing (decreasing) for $\microprobability < 1/2$ ($\microprobability > 1/2$). According to Equation \eqref{eq:majorityruledecoding}, $\macroprobability(\microprobability,\dimension)$ is a monotonically increasing function with respect to $\microprobability$ and $\macroprobability(1/2,\dimension)=1/2$, thus we prove that $\mathbb{S}(\macroprobability,\dimension)$ is also monotonically increasing (decreasing) for $\macroprobability < 1/2$ ($\macroprobability > 1/2$).

\section{Detailed Derivation of Optimal Erasure Protocol of Arrhenius-Rates Unit Model}\label{sec:optimalprotocolarrheniusderivation}

For Arrhenius rates, the master Equation \eqref{eq:masterequationarrhenius} can be recast as follows
\begin{align}
\invtemperature \Delta \epsilon=- \ln \left( \frac{\dot{\microprobability} + \microprobability}{1-\microprobability} \right) .
\label{eq:masterequationarrheniusrewritten}
\end{align}
The heat dissipated by a microscopic Arrhenius unit is given by the functional
\begin{align}
\invtemperature \microheat = \invtemperature \int_{0}^{\tau} \Delta \epsilon \dot{\microprobability} \d t = \invtemperature \int_{\microprobability_i}^{\microprobability_f} \Delta \epsilon \, \d \microprobability = \int_{0}^{\tau}\mathcal{L}(\microprobability,\dot{\microprobability}) \d t,
\label{eq:heatfunctional}
\end{align}
with the explicitly time-independent Lagrangian 
\begin{align}
\mathcal{L}(\microprobability,\dot{\microprobability}) \equiv - \dot{\microprobability} \ln \left( \frac{\dot{\microprobability} + \microprobability}{1 - \microprobability} \right) .\label{eq:lagrangian}
\end{align}
The minimization of the heat functional amounts to solving the Euler-Lagrange equation
\begin{align}
\mathcal{L} - \dot{\microprobability} \frac{\partial \mathcal{L}}{\partial \dot{\microprobability} } = K  ,
\label{eq:lagrangeequation}
\end{align}
that admits the solutions
\begin{align}
\dot{\microprobability}_1 = \frac{K - \sqrt{K^{2} + 4 K \microprobability}}{2}, \quad 
\dot{\microprobability}_2 = \frac{K + \sqrt{K^{2} + 4 K \microprobability}}{2} ,\label{eq:lagrangeequationsolution}
\end{align}
where $K$ is constant resulting from the time-integration of the Euler-Lagrange equation.

Since we consider the erasure branch from the initial state $\microprobability_i = 1/2$ to the final one $\microprobability_f \leq p_{i}$, we restrict to the solution $\dot{p}_1$. This ordinary differential equation yields the following explicit expression of the erasure duration $\tau$
\begin{align}
\tau=\int _{\microprobability_{i}}^{\microprobability_{f}} \frac{1}{\dot{\microprobability}} \d \microprobability = F(\microprobability_{f}) - F(\microprobability_{i}) ,
\label{eq:erasuretimeexpression}
\end{align}
where we defined the function $F(\microprobability)$ as
\begin{align} \label{eq:erasuretimeexpressionfunctiondefinition}
F(\microprobability) = - \sqrt{1 + \frac{4 \microprobability}{K}} - \ln \left( \sqrt{1 + \frac{4 \microprobability}{K}} - 1 \right) .
\end{align}
Substituting Equation \eqref{eq:lagrangeequationsolution} into Equation \eqref{eq:heatfunctional}, we obtain the following expression for the dissipated~heat
\begin{align}
\invtemperature \microheat = \invtemperature \int_{\microprobability_i}^{\microprobability_f} \Delta \epsilon \, \d \microprobability =  G(\microprobability_{f}) - G(\microprobability_{i})  ,
\label{eq:heatfunctionalsolution}
\end{align}
where we defined the function $G(\microprobability)$ as
\begin{align} 
G(p) = \frac{1}{2} \sqrt{K^{2}+4K \microprobability} - \ln{(1-\microprobability)} - \microprobability \ln \left( \frac{K + 2\microprobability - \sqrt{K^{2}+4K \microprobability}}{2(1-\microprobability)} \right) .
\end{align}

\bibliography{bibliography}

\end{document}